**Multimethod geophysical modelling for granite-related tungsten exploration: example of the Puy-les-Vignes/ Saint-Goussaud district (Limousin, France)**

*Dubreuil G. (1), Harlaux M. (1), Martelet G. (1), Vic G. (1), Bernard J. (1), Reninger P.-A. (1), Raingeard A. (1), Dubois F. (1), Peyrefitte A. (1)*

*(1) BRGM, 3 Av. Claude Guillemin, 45100 Orléans, France*

**Introduction**

In the last decade, the EU became aware of its high dependency on mining countries for mineral resources used by several strategic industries. To face it, the EU's commission created in 2023 the Critical Raw Materials Act aiming to secure a sustainable supply of mineral resources to help Europe to realize its 2030 climate and digital objectives. In this context, France launched new airborne geophysical surveys on former mining districts like the Limousin region in the northwestern French Massif Central. This area has produced significant amounts of gold, uranium, tungsten and tin ores.

The Limousin region is related to the European Variscan orogeny. It is mostly composed of medium- to high-grade gneisses, micaschists and amphibolites, which were intruded by several biotite-cordierite granites and two-mica leucogranites.

Tungsten and tin mineralization is spatially and temporally associated with these latter evolved leucogranites, and occurs as veins and stockworks within granitic cupolas and their metamorphic host rocks. The Puy-les-Vignes deposit, located in the St-Léonard-de-Noblat area, is the most important tungsten deposit in the Limousin region, and consists of a wolframite-mineralized hydrothermal breccia pipe hosted by migmatitic paragneisses. Located 10 km southwest of the Auriat granite, it has no clear spatial association with any known leucogranite cupola [Harlaux et al., 2021].

Additionally, this ore deposit type has no clear geophysical footprint, due in part to the lack of magnetic body (i.e., massive sulphides and/or magnetite) and its limited spatial extent, therefore challenging the exploration of hidden deposits. It is thus necessary to develop new exploration tools based on favourable environments, such as the presence of evolved granitic cupolas, permeable structures and hydrothermal alteration zones associated [Fang et al., 2015]. To do so, geological 3D modelling constrained by multimethod geophysical data can help to understand the architecture of the basement and enhance exploration targeting. We present the results of a 36 x 22 x 7 km geological model coupled with multimethod geophysical data, which has been built on the Saint-Goussaud – Auriat area centred on the Puy-les-Vignes district.

**Data compilation and Methodology**

Prior to modelling, geological and geophysical data have been compiled and processed. First, surface geometry constraints such as lithological limits of the main geological units and regional structures have been collected from published 1:50,000-scale geological maps (Figure 1-A). Fifty-five foliation measure points have also been integrated in the geological model. Despite limited outcropping conditions, bedrock geology is relatively well known near surface, but the extension of geological units at depth is uncertain. Three deep boreholes (>500 m) inside the Auriat and Saint-Goussaud leucogranites have been integrated in the model. However, none of them reached the granite basement, giving constraints on their minimal thicknesses, of about 1 km.

Thanks to the previous national mineral survey, W-Sn-mineralized occurrences and stream-sediment geochemical anomalies (>60 ppm W) have been reported in proximity to the granitic plutons. They have also been added to the workflow to define areas of interest.

To tackle the lack of constraints at depth, geophysical data have been integrated. The area is densely covered with 1 point/km² gravimetric dataset, extracted from gravimetric surveys on the metropolitan French territory. In 1999, airborne magnetic and gamma-spectrometric data have been acquired in the region. In 2022, a helicopter-borne magnetic and electromagnetic (TDEM) geophysical survey has been carried out to complete the geophysical cover (Figure 1-B). This latter method adds depth constraints on imaged bodies and allows investigating deeper than 1 km in the best conditions. Survey lines are separated by 400 m and tie lines by 4 km. The ground clearance was maintained at 50 m to drape the topography and maximize the investigation depth.

Petrophysical properties such as magnetic susceptibility and density have been compiled from similar lithologies elsewhere in the French Massif Central. This dataset has been used to determine petrophysical parameters in the geophysical modelling [Baptiste et al., 2016; Gebelin et al., 2004].

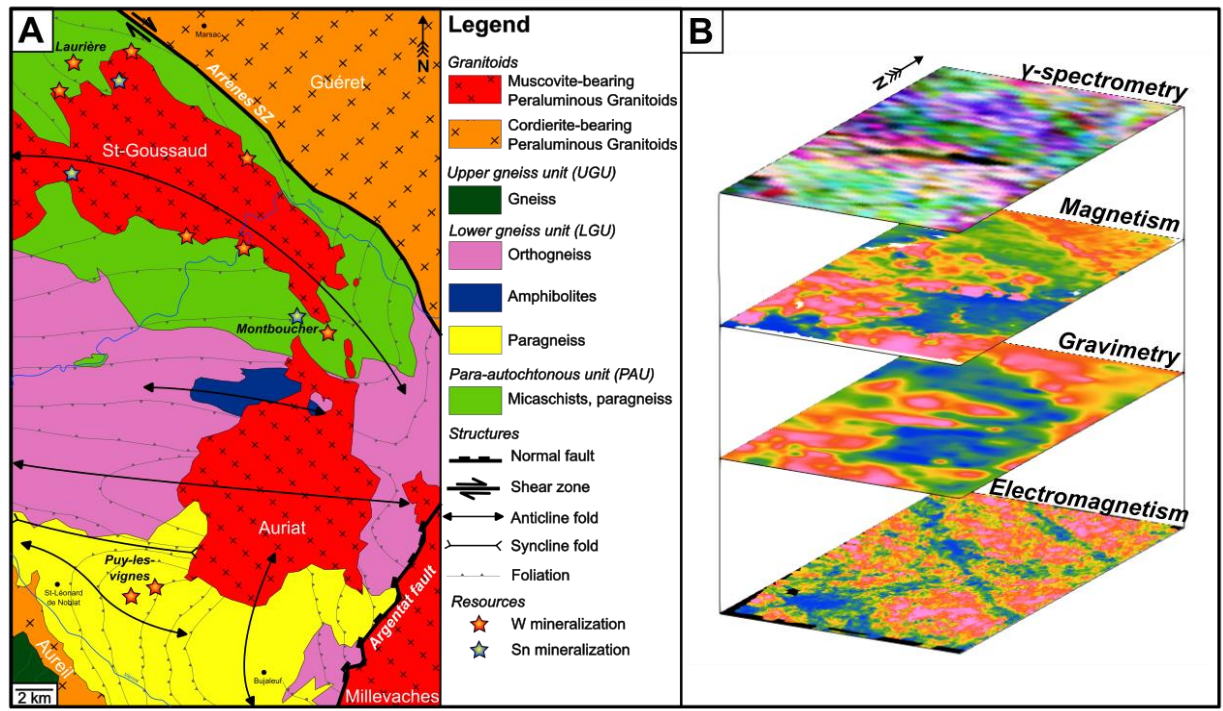

*Figure 1: (A) Simplified geological map of the Puy-les-Vignes district showing known W-Sn occurrences. (B) Airborne and ground geophysical datasets available in the area (gamma-spectrometry, magnetism, gravimetry, and electromagnetism)*

After compilation, geophysical data have been processed in the workflow. On one hand, helicopter-borne magnetic data have been manually edited to remove anthropic noise and every non-geological magnetic source. On the other hand, airborne electromagnetism (AEM) data have been automatically filtered to remove the electromagnetic noise mostly generated by power lines and other facilities. Classical processing has then been applied on both magnetic and gravimetric data. Magnetic data have been reduced to magnetic poles and the current main field effect has been removed to work with anomalies. Likewise, a regional trend has been subtracted from gravimetric data with a 30 km high-pass filter.

To be able to interpret magnetic decays from the AEM survey, they needed to be processed with an inversion algorithm. Using a smooth inversion style, data have been converted into a 3D resistivity block, bringing a resistivity value up to 1 km deep in some areas.

The first and easiest step to understand the basement architecture was to interpret directly processed geophysical grids. One of the most relevant work was to draw lineaments on AEM grids at different depths using the QGIS software. Lineaments were defined where it was possible to draw straight lines at the limit between two relatively resistive and conductive bodies. In the process, resistivity clusters have been defined on resistivity grid to distinguish resistive (> 1500 $\Omega$.m) from conductive zones (< 300 $\Omega$.m).

Besides, magnetic and gravimetric data have been exploited using GM-SYS software. In order to build vertical cross-sections by direct modelling constrained with geophysics, measured data have been compared to the modelled signal. In this case, it was possible to draw geophysical body interfaces while attributing them a density and magnetic susceptibility value. Moreover, corresponding resistivity cross-sections were plotted in the background to extrapolate interfaces from surface to depth. Interfaces were then iteratively moved to reduce the misfit between measured and modelled anomalies. In that case study, five cross-sections have been built: three are oriented SW to NE crossed by SE-NW and N-S last ones. At each crossing point, interfaces have been adjusted to be at the same depth.

Finally, geological data such as surficial body limits, regional structures and foliation direction-dip data have been implemented into the 3D modelling Geomodeller software. Geophysical cross-sections have then been relocated and digitalized into the geological 3D model, following interfaces interpolation.

## Results & Discussion

Mineralized occurrences and geochemistry anomalies compilation has allowed to highlight promising districts for exploration of granite-related W-Sn mineralization. One of them is located around the Puy-les-Vignes breccia pipe, based on a W-As stream-sediment anomaly. At the northwestern Saint-Goussaud termination, wolframite-bearing quartz veins have been reported nearby the Laurière locality as well as tin geochemical anomalies. A third location around the Montboucher village is characterized by W mineralization, stream sediment anomalies and its proximity with both Auriat and Saint-Goussaud leucogranites.

Lineament analysis from electromagnetic data has led to a 3D lineament network. The most frequent direction is oriented N135°. Then comes a perpendicular cluster, mainly oriented N40°. One last cluster is oriented along the North to South axis. These directions echo known regional structures in the area. The Arrènes shear zone which can be observed as a conductive corridor is oriented in the same direction as the main lineament cluster. Another regional corridor parallel to the Arrènes structure has been identified in the area, cutting the Auriat granite in two and joining N110° structures at the west of the Saint-Goussaud granite. This corridor could be interpreted as an unknown regional structure, named 'Sauviat fault' in Figure 2. Overall, the N135° direction does not match mapped faults in the area, which are dominantly oriented N-S or NE-SW. Instead, this trend could be interpreted as a hypothetical fault, which was not recognized.

Working on resistivity grids from the AEM survey, anomalous low-resistivity zones are been detected. They are defined as vertical conductors in the first 200m investigated. The largest anomaly is located on the western part of the Puy-les-Vignes deposit. Others occur close to granites or along regional structures such as the 'Sauviat' fault. Except the Guéret granite northeast of the study area, every anomaly is associated with W-Sn geochemical anomalies. These vertical conductors could represent deep hydrothermal alteration corridors, associated with vertical permeable conduits.

Besides, multimethod geophysical coupling has allowed modelling the geometry of the Saint-Goussaud and Auriat granites. As shown in Figure 2, some granite parts extend further into their host rocks. For instance, the Saint-Goussaud granite slightly deepens under the Laurière and Montboucher districts. These areas may represent potential granitic cupolas where mineralization could be localized. According to the gravimetric signal, the Auriat granite plunges and thickens to the SW into paragneisses. In this 3D model, it is not obviously related with the Puy-les-Vignes district. However, a limit interpreted as the Auriat extension at depth can be extracted from the tilt derivative operator applied on complete Bouguer anomaly. This limit extends from the Auriat pluton to the Puy-les-Vignes district, which raises questions about the presence of an unexposed granite body at depth. However, one must be careful while interpreting these extensions since granites geometry is fiercely dependent on the five geophysical cross-sections position and data coverage. Indeed, the area is well covered with gravimetric data in the South but points get scarce around the Saint-Goussaud granite.

## Conclusions

This work offers a new approach for granite-related W-Sn exploration, using multimethod geophysical and geological modelling. The obtained results allowed proposing the existence of unexposed granitic cupolas around the Saint-Goussaud and Auriat granites, which could be responsible for the numerous W-Sn occurrences and geochemical anomalies at the surface. Moreover, granitic cupolas can be visualized thanks to resistivity cross-sections from the AEM data. This method has also imaged anomalous low-resistivity zones, often located on N130° faulted corridors and sometimes related to W-Sn-(As) geochemical anomalies. All these combined elements could have implications for exploration targeting of granite-related W-Sn deposits at a regional scale.

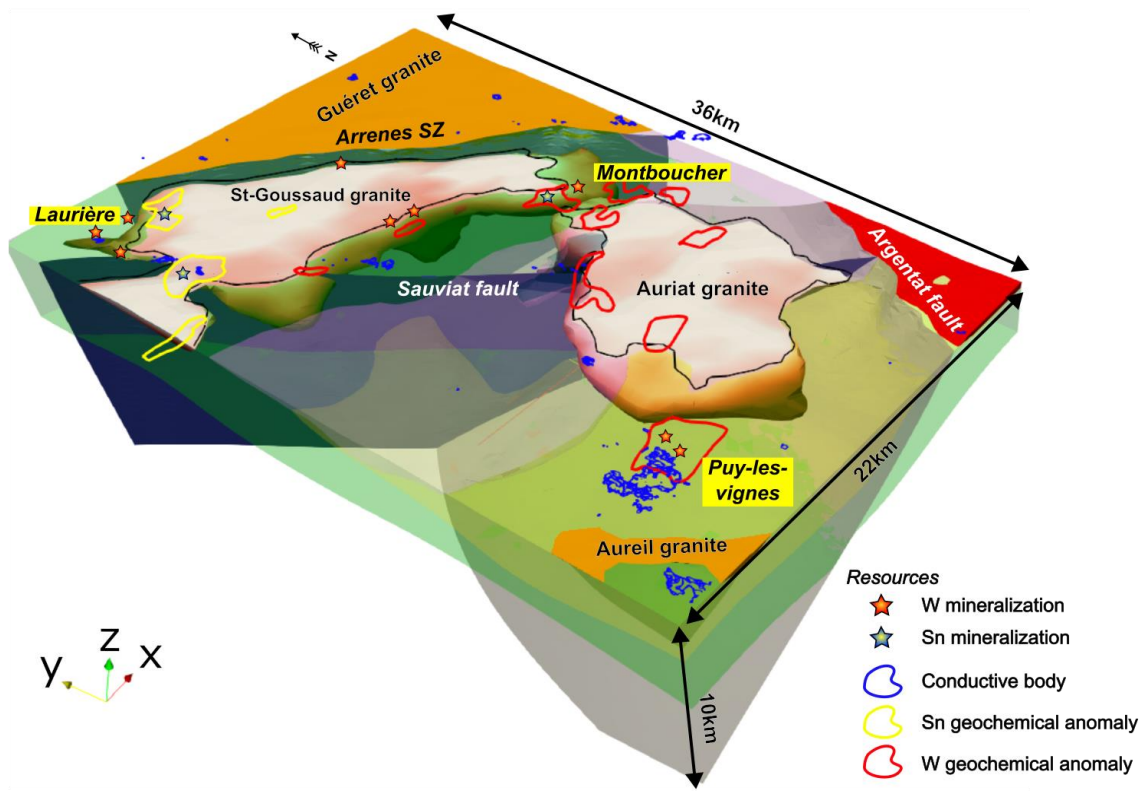

*Figure 2: Regional 3D geological model of the Puy-les-Vignes / St-Goussaud districts. Prospective areas for W exploration are defined near a granitic cupola or/and an anomalous low resistivity zone, in a faulted area and matching a W-(Sn) geochemical anomaly.*

**Acknowledgements**
This work was supported by funds from the French Ministry of Ecological Transition and the BRGM.